\documentclass{amsart}
\usepackage{amsfonts}
\usepackage{amsmath}
\usepackage{amssymb}

\theoremstyle{plain}
\newtheorem{theorem}{Theorem}[section]
\newtheorem{lemma}[theorem]{Lemma}
\newtheorem{proposition}[theorem]{Proposition}
\theoremstyle{definition}

\theoremstyle{remark}
\newtheorem{remark}[theorem]{Remark}
\numberwithin{equation}{section}

\input{tcilatex}

\begin{document}
\title[The Lorentz condition problem]{The electromagnetic Lorentz condition
problem and symplectic properties of Maxwell and Yang-Mills type dynamical
systems}
\author{N.N. Bogolubov (jr.)}
\address{V.A. Steklov Mathematical Institute of RAS, Moscow, Russian
Federation\\
and\\
The Abdus Salam International Centre of Theoretical Physics, Trieste, Italy}
\email{nikolai\_bogolubov@hotmail.com}
\author{A.K. Prykarpatsky}
\address{The AGH University of Science and Technology, Department of Applied
Mathematics, Krakow 30059 Poland\\
and\\
Ivan Franko Pedagogical State University, Drohobych, Lviv region, Ukraine \ }
\email{pryk.anat@ua.fm, prykanat@cybergal.com}
\author{U. Taneri}
\address{Department of Applied Mathematics and Computer Science, Eastern
Mediterranean University EMU, Famagusta, North Cyprus\\
and\\
Kyrenia American University GAU, Institute of Graduate Studies, Kyrenia,
North Cyprus }
\email{ufuk.taneri@gmail.com}
\author{ Y.A. Prykarpatsky}
\address{Ivan Franko Pedagogical State University, Drohobych, Lviv region,
Ukraine \ and \\
the Pedagogical University, Krakow, Poland}
\email{yarpry@gmail.com}
\date{present}

\begin{abstract}
Symplectic structures associated to connection forms on certain types of
principal fiber bundles are constructed via analysis of reduced geometric
structures on fibered manifolds invariant under naturally related symmetry
groups.This approach is then applied to nonstandard Hamiltonian analysis of
of dynamical systems of Maxwell and Yang-Mills type. A symplectic reduction
theory of the classical Maxwell equations is formulated so as to naturally
include the Lorentz condition (ensuring the existence of electromagnetic
waves), thereby solving the well known Dirac -Fock - Podolsky problem.
Symplectically reduced Poissonian structures and the related classical
minimal interaction principle for the Yang-Mills equations are also
considered.
\end{abstract}

\maketitle

\section{Introduction}

When investigating dynamical systems, which are invariant under symmetry
group actions, on canonical symplectic manifolds, additional mathematical
structures often arise. Analysis of these structures almost invariably
produces important dynamical insights about the systems. For example, the
Cartan connection on an associated principal fiber bundle leads to a more
detailed understanding of the reductions of the dynamical system on
invariant submanifolds and quotient manifolds.

Problems related to the investigation of properties of reduced dynamical
systems on symplectic manifolds were studied, e.g., in \cite%
{2AM,3PM,3HPP,10Pr,10PS}, where the relationship between a symplectic
structure on the reduced space and the connection on a principal fiber
bundle was explicitly formulated. Other aspects of dynamical systems related
to properties of reduced symplectic structures were studied in \cite%
{4Ku,5Ra,6HK} where, in particular, the reduced symplectic structure was
completely described within the framework of the classical Dirac scheme, and
several applications to nonlinear (including celestial) dynamics were given.

It is well known \cite{BS,Th, Di,BPT,BP,Pa} that the Hamiltonian formulation
of Maxwell's electromagnetic field equations involves a very important
classical problem; namely, to intrinsically introduce the Lorentz condition,
which guarantees the wave structure of propagating quanta and the positivity
of energy. Unfortunately, in spite of extensive classical studies by Dirac,
Fock and Podolsky \cite{DFP}, the problem remains open. Consequently, the
Lorentz condition is usually imposed in modern electrodynamics as an
external constraint rather than arising naturally from the Hamiltonian (or
Lagrangian) theory. Moreover, it was shown by Pauli, Dirac, Bogolubov and
Shirkov and others \cite{BS,Pa,Di,BjD} that the quantum Lorentz condition is
incompatible with existing quantization approaches for electromagnetic field
theory, except in an average sense. These difficulties stimulated our study
of this problem using symplectic reduction theory, which allows a systematic
introduction of the external charge and current conditions into the
Hamiltonian formalism, and actually leads to the solution to the Lorentz
condition problem described herein.

Some applications of the method to Yang-Mills type equations interacting
with a point charged particle are presented. In particular, by analyzing
reduced geometric structures on fibered manifolds invariant under the action
of a symmetry group, we construct the symplectic structures associated with
connection forms on suitable principal fiber bundles. We begin with a brief
description of the mathematical preliminaries of the related Poissonian
structures on the corresponding reduced symplectic manifolds, which are
often used \cite{2AM,9MW,8Kup} in various problems of dynamics in modern
mathematical physics. These methods are then applied to studying the
nonstandard Hamiltonian properties of Maxwell and Yang-Mills type dynamical
systems.

Our main contribution here is a novel formulation of a symplectic reduction
theory for the classical Maxwell electromagnetic field equations that
provides a means of naturally including the Lorentz condition (ensuring \cite%
{BS,BjD} the existence of electromagnetic waves) in the associated
Hamiltonian structure - thereby solving the Dirac-Fock-Podolsky \cite{DFP}
problem mentioned above. In addition, we also use our symplectic reduction
theory to investigate the Poissonian structures and the classical minimal
interaction principle related to Yang-Mills equations.

\section{Symplectic structures and reduction on manifolds: preliminaries}

In this section, we shall outline the basic elements of symplectic
structures and reduction on manifolds employed in the sequel.

\subsection{Symplectic reduction on cotangent fiber bundles with symmetry}

Consider an $n$-dimensional smooth manifold $M$ and the cotangent vector
fiber bundle $T^{\ast }(M).$ We equip (see \cite{Go}, Chapter VII) the
cotangent space $T^{\ast }(M)$ with the canonical Liouville 1-form $\lambda
(\alpha ^{(1)}):=$ $pr_{M}^{\ast }\alpha ^{(1)}\in \Lambda ^{1}(T^{\ast
}(M)),$ where $pr_{M}:T^{\ast }(M)\rightarrow M$ is the canonical projection
and%
\begin{equation}
\alpha ^{(1)}(u)=\sum_{j=1}^{n}v_{j}du^{j},  \label{1}
\end{equation}%
where $(u,v)\in T^{\ast }(M)$ are the corresponding canonical local
coordinates on $T^{\ast }(M).$ Thus, any group of diffeomorphisms of the
manifold $M$ naturally lifted to the fiber bundle $T^{\ast }(M)$ preserves
the invariance of the canonical 1-form $\lambda (\alpha ^{(1)})\in \Lambda
^{1}(T^{\ast }(M)).$ In particular, if a smooth action of a Lie group $G$ is
given on the manifold $M,$ then every element $a\in \mathcal{G},$ where $%
\mathcal{G}$ is the Lie algebra of the Lie group $G,$ generates the vector
field $k_{a}\in T(M)$ in a natural manner. Furthermore, since the group
action on $M,$ i.e., 
\begin{equation}
\varphi :G\times M\rightarrow M,  \label{2}
\end{equation}%
generates a diffeomorphism $\ \varphi _{g}\in Diff$ $M$ for every element $%
g\in G,$ this diffeomorphism lifts naturally to the corresponding
diffeomorphism $\varphi _{g}^{\ast }$ $\in Diff$ $T^{\ast }(M)$ of the
cotangent fiber bundle $T^{\ast }(M),$ which also leaves \ the canonical
1-form $pr_{M}^{\ast }\alpha ^{(1)}\in \Lambda ^{1}(T^{\ast }(M))$
invariant; namely,

\begin{equation}
\varphi _{g}^{\ast }\lambda (\alpha ^{(1)})=\lambda (\alpha ^{(1)})
\label{3}
\end{equation}%
holds \cite{2AM,Go,3PM} for every 1-form $\alpha ^{(1)}\in \Lambda ^{1}(M).$
Thus, we can define on $T^{\ast }(M)$ the corresponding vector field $%
K_{a}:T^{\ast }(M)\rightarrow T(T^{\ast }(M))$ for every element $a\in 
\mathcal{G}.$ Then condition (\ref{3}) can be rewritten in the following
form for all $a\in \mathcal{G}:$

\begin{equation*}
L_{K_{a}}\cdot pr_{M}^{\ast }\alpha ^{(1)}=pr_{M}^{\ast }\cdot
L_{k_{a}}\alpha ^{(1)}=0,
\end{equation*}%
where $L_{K_{a}}$ and $L_{k_{a}}$ are the ordinary Lie derivatives on $%
\Lambda ^{1}(T^{\ast }(M))$ and $\Lambda ^{1}(M),$ respectively.

The canonical symplectic structure on $T^{\ast }(M)$ is defined as

\begin{equation}
\omega ^{(2)}:=d\lambda (\alpha ^{(1)})  \label{4}
\end{equation}%
and is invariant, i.e., $L_{K_{a}}\omega ^{(2)}=0$ for all $a\in \mathcal{G}%
. $

For any smooth function $H\in D(T^{\ast }(M)),$ \ a Hamiltonian vector field 
$K_{H}:T^{\ast }(M)\rightarrow T(T^{\ast }(M))$ \ such that

\begin{equation}
i_{K_{H}}\omega ^{(2)}=-dH  \label{5}
\end{equation}%
is defined, and vice versa, because the symplectic 2-form (\ref{4}) is
nondegenerate. Using (\ref{5}) and (\ref{4}), we easily establish that the
Hamiltonian function $H:=H_{K}\in D(T^{\ast }(M))$ is given as $%
H_{K}=pr_{M}^{\ast }\alpha ^{(1)}(K_{H})=\alpha ^{(1)}(pr_{M}^{\ast
}K_{H})=\alpha ^{(1)}(k_{H}),$ where $k_{H}\in T(M)$ is the corresponding
vector field on the manifold $M,$ whose lift to the fiber bundle $T^{\ast
}(M)$ coincides with the vector field $K_{H}:T^{\ast }(M)\rightarrow
T(T^{\ast }(M)).$ For $K_{a}:T^{\ast }(M)\rightarrow T(T^{\ast }(M)),$ where 
$a\in \mathcal{G},$ \ it is easy to establish that the corresponding
Hamiltonian function $H_{a}=\alpha ^{(1)}(k_{a})=pr_{M}^{\ast }$ $\alpha
^{(1)}(K_{a})$ for $a\in \mathcal{G}$ defines \cite{2AM,3PM,3HPP} a linear
momentum \ mapping $l:T^{\ast }(M)\rightarrow \mathcal{G}^{\ast }$ according
to the rule

\begin{equation}
H_{a}:=<l,a>,  \label{6}
\end{equation}%
where $<$\textperiodcentered $,$\textperiodcentered $\ >$ \ is the
corresponding convolution on $\mathcal{G}^{\mathcal{\ast }}$ $\times 
\mathcal{G}.$ By virtue of definition (\ref{6}), the momentum mapping $%
l:T^{\ast }(M)\rightarrow \mathcal{G}^{\ast }$ is invariant under the action
of any invariant Hamiltonian vector field $K_{b}:T^{\ast }(M)\rightarrow
T(T^{\ast }(M))$ for any $b\in \mathcal{G}.$ Indeed, $%
L_{K_{b}}<l,a>=L_{K_{b}}H_{a}=-L_{K_{a}}H_{b}=0,$ because, by definition,
the Hamiltonian function $H_{b}\in D(T^{\ast }(M))$ is invariant under the
action of any vector field $K_{a}:T^{\ast }(M)\rightarrow T(T^{\ast }(M)),$ $%
a\in \mathcal{G}.$

We now fix a regular value of the momentum mapping $l(u,v)=\xi \in \mathcal{G%
}^{\ast }$ and consider the corresponding submanifold $\mathcal{M}_{\xi
}:=\{(u,v)\in T^{\ast }(M):l(u,v)=\xi \in \mathcal{G}^{\ast }\}.$ Owing to
definition (\ref{1}) and the invariance of the 1-form $pr_{M}^{\ast }$ $%
\alpha ^{(1)}\in \Lambda ^{1}(T^{\ast }(M))$ under the action of the Lie
group $G$ on $T^{\ast }(M),$ we have

\begin{eqnarray}
&<&l(g\circ (u,v)),a>=pr_{M}^{\ast }\alpha ^{(1)}(K_{a})(g\circ (u,v))= 
\notag \\
&=&pr_{M}^{\ast }\alpha ^{(1)}(K_{Ad_{g-1}a})(u,v):=  \label{7} \\
&=&<l(u,v),Ad_{g-1}a>=<Ad_{g-1}^{\ast }l(u,v),a>  \notag
\end{eqnarray}%
for any $g\in G$ and all $a\in \mathcal{G}$ and $(u,v)\in T^{\ast }(M).$ Now
it follows from (\ref{7}) that $l(g\circ (u,v))=Ad_{g^{-1}}^{\ast }l(u,v)$\
for every $g\in G$ and all $(u,v)\in T^{\ast }(M).$ This means that the
diagram%
\begin{equation*}
\begin{array}{ccc}
T^{\ast }(M) & \overset{l}{\rightarrow } & \mathcal{G}^{\mathcal{\ast }} \\ 
g\downarrow &  & \downarrow Ad_{g-1}^{\ast } \\ 
T^{\ast }(M) & \overset{l}{\rightarrow } & \mathcal{G}^{\mathcal{\ast }}%
\end{array}%
\end{equation*}%
is commutative for all elements $g\in G.$ The corresponding action $%
g:T^{\ast }(M)\rightarrow T^{\ast }(M)$ is called equivariant \cite{2AM,3PM}.

Let $\ G_{\xi }\subset G$ \ denote the stabilizer of a regular element $\xi
\in \mathcal{G}^{\ast }$ with respect to the related co-adjoint action. \ It
is obvious in this case that the action of the Lie subgroup $G_{\xi }$ on
the submanifold $\mathcal{M}_{\xi }\subset T^{\ast }(M)$ is naturally
defined; we assume that it is free and proper. Using this action on $%
\mathcal{M}_{\xi },$ we can define \cite{2AM,5Ra,6HK,7Mo,8Kup} a so-called
reduced space $\mathcal{\bar{M}}_{\xi }$ by taking the factor with respect
to the action of the subgroup $G_{\xi }$ on $\mathcal{M}_{\xi },$ i.e.,

\begin{equation}
\mathcal{\bar{M}}_{\xi }:=\mathcal{M}_{\xi }/G_{\xi }.  \label{8}
\end{equation}%
The quotient space (\ref{8}) induces a symplectic structure $\bar{\omega}%
_{\xi }^{(2)}\in \Lambda ^{2}(\mathcal{\bar{M}}_{\xi })$ on itself, which is
defined as follows:%
\begin{equation}
\bar{\omega}_{\xi }^{(2)}(\bar{\eta}_{1},\bar{\eta}_{2})=\omega _{\xi
}^{(2)}(\eta _{1},\eta _{2}),  \label{9}
\end{equation}%
where $\bar{\eta}_{1},\bar{\eta}_{2}\in T(\mathcal{\bar{M}}_{\xi })$ are
arbitrary vectors onto which vectors $\eta _{1},\eta _{2}\in T(\mathcal{M}%
_{\xi })$ are projected for at any point $(u_{\xi },v_{\xi })\in \mathcal{M}%
_{\xi }.$It follows from (\ref{8}) that this projection onto the point $\bar{%
\mu}_{\xi }\in \mathcal{\bar{M}}_{\xi }$ is unique .

Let $\pi _{\xi }:\mathcal{M}_{\xi }\rightarrow T^{\ast }(M)$ denote the
corresponding imbedding mapping into $T^{\ast }(M)$ and let $r_{\xi }:%
\mathcal{M}_{\xi }\mathcal{\rightarrow \bar{M}}_{\xi }$ be the corresponding
reduction to the space $\mathcal{\bar{M}}_{\xi }.$ Then relation (\ref{9})
can be rewritten in the form 
\begin{equation}
r_{\xi }^{\ast }\bar{\omega}_{\xi }^{(2)}=\pi _{\xi }^{\ast }\omega ^{(2)},
\label{10}
\end{equation}%
which is defined on vectors on the cotangent space $T^{\ast }(\mathcal{M}%
_{\xi }).$ To establish the symplecticity of the 2-form $\omega _{\xi
}^{(2)}\in \Lambda ^{2}(\mathcal{\bar{M}}_{\xi }),$ we use the corresponding
non-degeneracy of the Poisson bracket $\{$\textperiodcentered $,$%
\textperiodcentered $\}_{\xi }^{r}$ on $\mathcal{\bar{M}}_{\xi }.$ We use a
Dirac type construction for the calculation, defining functions on $\ 
\mathcal{\bar{M}}_{\xi }$ as certain $G_{\xi }$-invariant functions on the
submanifold $\mathcal{M}_{\xi }.$ Then one can calculate the Poisson bracket 
$\{$\textperiodcentered $,$\textperiodcentered $\}_{\xi }$ of such a
function corresponding to the symplectic structure (\ref{4}) as an ordinary
Poisson bracket on $T^{\ast }(M),$ arbitrarily extending these functions
from the submanifold $\mathcal{M}_{\xi }$ $\subset T^{\ast }(M)$ to a
neighborhood $U(\mathcal{M}_{\xi })\subset T^{\ast }(M).$ It is obvious that
two extensions of a given function to the neighborhood $U(\mathcal{M}_{\xi
}) $ of this type differ by a function that vanishes on the submanifold $%
\mathcal{M}_{\xi }\subset T^{\ast }(M).$ The difference between the
corresponding Hamiltonian fields of these two different extensions to $U(%
\mathcal{M}_{\xi })$ is completely controlled by the conditions of the
following lemma (see also \cite{2AM,3PM,6HK,5Ra,10Pr}).

\begin{lemma}
\label{Lm_2.1}Suppose that a function $f:U(\mathcal{M}_{\xi })\rightarrow 
\mathbb{R}$ is smooth and vanishes on $\mathcal{M}_{\xi }$ $\subset T^{\ast
}(M),$ i.e., $f|_{\mathcal{M}_{\xi }}=0.$ Then, at every point $(u_{\xi
},v_{\xi })\in \mathcal{M}_{\xi }$ the corresponding Hamiltonian vector
field $K_{f}\in T(U(\mathcal{M}_{\xi }))$ is tangent to the orbit $%
Or(G;(u_{\xi },v_{\xi })).$
\end{lemma}

As a corollary of Lemma \ref{Lm_2.1}, we obtain an algorithm for computing
the reduced Poisson bracket $\{$\textperiodcentered $,$\textperiodcentered $%
\}_{\xi }^{r}$ on the space $\mathcal{\bar{M}}_{\mathcal{\xi }}$ according
to definition (\ref{10}). Namely, we choose two functions defined on $%
\mathcal{M}_{\xi }$ and invariant under the action of the subgroup $G_{\xi }$
and arbitrarily smoothly extend them to a certain open domain $U(\mathcal{M}%
_{\xi })\subset T^{\ast }(M).$ Then we determine the corresponding
Hamiltonian vector fields on $T^{\ast }(M)$ and project them onto the space
tangent to $\mathcal{M}_{\xi },$ adding, if necessary, the corresponding
vectors tangent to the orbit $Or(G).$ It is easy to see that the projections
obtained depend on the chosen extensions to the domain $U(\mathcal{M}_{\xi
})\subset T^{\ast }(M).$ As a result, we establish that the reduced Poisson
bracket $\{$\textperiodcentered $,$\textperiodcentered $\}_{\xi }^{r}$ is
uniquely defined via the restriction of the initial Poisson bracket upon $%
\mathcal{M}_{\xi }$ $\subset T^{\ast }(M),$ and one can readily verify that
the submanifold $M_{\xi }\subset T^{\ast }(M)$ is defined by a collection of
relations of the type%
\begin{equation}
H_{a_{s}}=\xi _{s},\text{ \ \ \ \ \ \ \ \ }\xi _{s}:=<\xi ,a_{s}>,
\label{11}
\end{equation}%
where $a_{s}\in \mathcal{G},s=\overline{1,dimG},$ is a certain basis of the
Lie algebra $\mathcal{G}.$ By virtue of the nondegeneracy of the restriction
and the functional independence of the basis functions (\ref{11}), it is
obvious that the reduced Poisson bracket $\{$\textperiodcentered $,$%
\textperiodcentered $\}_{\xi }^{r}$ is \cite{2AM,3PM,5Ra} nondegenerate on $%
\mathcal{\bar{M}}_{\xi }.$ \ Consequently, we establish that the dimension
of the reduced space $\mathcal{\bar{M}}_{\xi }$ is even. Taking into account
that the element $\xi \in \mathcal{G}^{\mathcal{\ast }}$ is regular and the
dimension of the Lie algebra of the stabilizer $\mathcal{G}_{\xi }$ is equal
to $dim$ $G_{\xi },$ we easily establish that $dim$ $\mathcal{\bar{M}}_{\xi
}=$ $dim$ $T^{\ast }(M)-2dim$ $\mathcal{G}_{\xi }.$ Since, by construction, $%
dim$ $T^{\ast }(M)=2n,$ we conclude that the dimension of the reduced space $%
\mathcal{\bar{M}}_{\xi }$ is even.

In order completely verify the correctness of the algorithm, it is necessary
to establish the existence of the corresponding projections of Hamiltonian
vector fields onto the tangent space $T(\mathcal{M}_{\xi }).$ The following
result \cite{10PS} solves this problem.

\begin{theorem}
\label{Tm_2.1}At every point $(u_{\xi },v_{\xi })\in \mathcal{M}_{\xi },$
one can choose a vector $V_{f}\in T(Or(G))$ such that $K_{f}(u_{\xi },v_{\xi
})$ $+V_{f}(u_{\xi },v_{\xi })\in T_{(u_{\xi },v_{\xi })}(\mathcal{M}_{\xi
}).$ Furthermore, the vector $V_{f}\in T(Or(G))$ is uniquely determined up
to a vector tangent to the orbit $Or(G_{\xi }).$
\end{theorem}

Now assume that two functions $f_{1},f_{2}\in D(\mathcal{M}_{\xi })$ are $%
G_{\xi }$-invariant. Then their reduced Poisson bracket $\{f_{1},f_{2}\}_{%
\xi }^{r}$ on $\mathcal{\bar{M}}_{\xi }$ is defined according to the rule: 
\begin{equation}
\{f_{1},f_{2}\}_{\xi }^{r}:=-\omega
^{(2)}(K_{f_{1}}+V_{f_{1}},K_{f_{2}}+V_{f_{2}})=\{f_{1},f_{2}\}+\omega
^{(2)}(V_{f_{1}},V_{f_{2}}),  \label{17}
\end{equation}%
where we have used the following identities on $\mathcal{M}_{\xi }$ $\subset
T^{\ast }(M):$ 
\begin{equation}
\omega ^{(2)}(K_{f_{1}}+V_{f_{1}},V_{f_{2}})=0=\omega
^{(2)}(K_{f_{2}}+V_{f_{2}},V_{f_{1}}),  \label{17aa}
\end{equation}%
which follow immediately from 
\begin{equation}
\omega ^{(2)}(K_{f}+V_{f},K_{a})=0  \label{17a}
\end{equation}%
for all $a\in \mathcal{G}_{\xi }$ and $f\in D(\mathcal{M}_{\xi })$ \ on $%
\mathcal{M}_{\xi }.$ With regard to (\ref{17aa}), relation (\ref{17}) takes
the form 
\begin{equation}
\{f_{1},f_{2}\}_{\xi }^{r}=\{f_{1},f_{2}\}+\frac{1}{2}%
(V_{f_{1}}f_{2}-V_{f_{2}}f_{1}),  \label{18}
\end{equation}%
for arbitrary smooth extensions $f_{1},f_{2}\in D(\mathcal{M}_{\xi })$ of $%
G_{\xi }$-invariant functions, as defined above on the domain $U(\mathcal{M}%
_{\xi }).$ Thus, as a consequence of (\ref{Tm_2.1}), one has the following 
\cite{2AM,Di,3PM} \ theorem of Dirac type.

\begin{theorem}
\label{Tm_2.2}The reduced Poisson bracket of two functions on the quotient
space $\mathcal{\bar{M}}_{\xi }$ $=$ $\mathcal{M}_{\xi }/G_{\xi }$ is
determined with the use of arbitrary smooth extensions of them to functions
on an open neighborhood $U(\mathcal{M}_{\xi })$ according to the Dirac-type
formula (\ref{18}).
\end{theorem}

\subsection{Symplectic reduction on principal fiber bundles with a connection%
}

We begin by reviewing reduction theory for Hamiltonian systems \ with
symmetry on principle fiber bundles. As the material is partially available
in \cite{1GS,4Ku}, we shall provide only a sketch here using notation that
is to be employed in the sequel.

Let $G$ denote a Lie group with the unity element $e\in G$ and $\mathcal{G}$ 
$\simeq T_{e}(G)$ be its Lie algebra$.$ Consider a principal fiber bundle $%
\pi :(M,\varphi )\rightarrow N$ with the structure group $G$ and base
manifold $N,$ on which the Lie group $G$ acts via a mapping $\ \varphi
:M\times G\rightarrow M.$ In particular, for each $g\in G$ there is a group
diffeomorphism $\varphi _{g}:M\rightarrow M,$ generating for any fixed $u\in
M$ the following induced mapping: $\hat{u}:G\rightarrow M,$ where 
\begin{equation}
\hat{u}(g)=\varphi _{g}(u).  \label{0.1}
\end{equation}

This mapping induces a connection $\Gamma (${$\mathcal{A}$}$)$ on the
principal fiber bundle $\pi :(M,\varphi )\rightarrow N$, where the morphism {%
$\mathcal{A}$}$:(T(M),\varphi _{g\ast })\rightarrow (\mathcal{G}%
,Ad_{g^{-1}}),$ such that for each $u\in M$ a mapping $\mathcal{A}%
(u):T_{u}(M)\rightarrow \mathcal{G}$ is a left inverse of the mapping \ $%
\hat{u}_{\ast }(e):\mathcal{G}\rightarrow T_{u}(M),$ that is 
\begin{equation}
\mathit{\mathcal{A}}(u)\hat{u}_{\ast }(e)=1.  \label{0.2}
\end{equation}%
As usual, we denote by $\varphi _{g}^{\ast }:T^{\ast }(M)\rightarrow T^{\ast
}(M)$ the corresponding lift of the mapping \ $\varphi _{g}:M\rightarrow M$
\ at any $g\in G.$ If $\alpha ^{(1)}\in \Lambda ^{1}(M)$ is the canonical $G$
- invariant 1-form on $\ M,$ the canonical symplectic structure $\omega
^{(2)}\in \Lambda ^{2}(T^{\ast }(M)),$ given by the expression 
\begin{equation}
\omega ^{(2)}:=d\lambda (\alpha ^{(1)})=d\text{ }pr_{M}^{\ast }\alpha ^{(1)},
\label{0.3}
\end{equation}%
generates the corresponding momentum mapping $l:T^{\ast }(M)\rightarrow 
\mathcal{G}^{\ast },$ where 
\begin{equation}
l\cdot \alpha ^{(1)}(u)=\hat{u}^{\ast }(e)\alpha ^{(1)}(u)  \label{0.4}
\end{equation}%
for all $u\in M.$ We remark here that the principal fiber \ bundle structure 
$\pi :(M,\varphi )\rightarrow N$ \ entails in part the exactness of the
following sequences of mappings: 
\begin{equation}
0\rightarrow \mathcal{G}\overset{\hat{u}_{\ast }(e)}{\rightarrow }T_{u}(M)%
\overset{\pi _{\ast }(u)}{\rightarrow }T_{\pi (u)}(N)\rightarrow 0,
\label{0.5}
\end{equation}%
that is 
\begin{equation}
\pi _{\ast }(u)\hat{u}_{\ast }(e)=0=\hat{u}^{\ast }(e)\pi ^{\ast }(u)
\label{0.6}
\end{equation}%
for all $u\in M.$ Combining (\ref{0.6}) with (\ref{0.2}) and (\ref{0.4}),
one obtains the embedding: 
\begin{equation}
\lbrack 1-\mathcal{A}^{\ast }(u)\hat{u}^{\ast }(e)]\alpha ^{(1)}(u)\in \text{%
range }\pi ^{\ast }(u)  \label{0.7}
\end{equation}%
for the canonical 1-form $\alpha ^{(1)}\in \Lambda ^{1}(M)$ at $u\in M.$ The
expression (\ref{0.7}) means of course, that 
\begin{equation}
\hat{u}^{\ast }(e)[1-\mathcal{A}^{\ast }(u)\hat{u}^{\ast }(e)]\alpha
^{(1)}(u)=0  \label{0.8}
\end{equation}%
for all $u\in M.$ As the mapping \ $\pi ^{\ast }(u):T^{\ast }(N)\rightarrow
T^{\ast }(M)$ \ is injective for each $u\in M$ , it has the unique inverse
mapping \ $\ (p^{\ast }(u))^{-1}$defined on its image \ $\pi ^{\ast
}(u)T_{\pi (u)}^{\ast }(N)\subset T_{u}^{\ast }(M).$ Whence, \ for each $%
u\in M$ one can define a morphism $\pi _{\mathcal{A}}:(T^{\ast }(M),\varphi
_{g}^{\ast })\rightarrow T^{\ast }(N)$ as 
\begin{equation}
\pi _{\mathcal{A}}(u):\alpha ^{(1)}(u)\rightarrow (\pi ^{\ast }(u))^{-1}[1-%
\mathcal{A}^{\ast }(u)\hat{u}^{\ast }(e)]\alpha ^{(1)}(u).  \label{0.9}
\end{equation}%
It is easy to check using (\ref{0.9}) that the diagram 
\begin{equation}
\begin{array}{ccc}
T^{\ast }(M) & \overset{\pi ^{\ast }}{\leftarrow } & T^{\ast }(N) \\ 
\left. pr_{M}\right\downarrow  &  & \left\downarrow pr_{N}\right.  \\ 
M & \overset{\pi }{\rightarrow } & N%
\end{array}
\label{0.10}
\end{equation}%
is commutative.

Now suppose an element $\xi \in \mathcal{G}^{\ast }$ be $G$-invariant, that
is $\ \ Ad_{g^{-1}}^{\ast }\xi =\xi $ for all $\ g\in G.$ Let \ $\pi _{%
\mathcal{A}}^{\xi }$ \ denote the restriction of the mapping (\ref{0.9})
upon the subset $\mathcal{M}_{\xi }:=l^{-1}(\xi )\in T^{\ast }(M),$ that is
\ \ $\pi _{\mathcal{A}}^{\xi }:\mathcal{M}_{\xi }\rightarrow T^{\ast }(N),$
where for all $u\in M$%
\begin{equation}
\pi _{\mathcal{A}}^{\xi }(u):l^{-1}(\xi )\rightarrow (\pi ^{\ast
}(u))^{-1}[1-\mathcal{A}^{\ast }(u)\hat{u}^{\ast }(e)]l^{-1}(\xi ).
\label{0.11}
\end{equation}%
The structure of the reduced phase space {$\mathcal{\bar{M}}_{\xi }:=$}$%
l^{-1}(\xi )/G$ \ can now be characterized by means of the following lemma.

\begin{lemma}
{\label{lem_01} {The mapping }}${\pi }${$_{\mathcal{A}}^{\xi }(u):\mathcal{M}%
_{\xi }\rightarrow T^{\ast }(N),$ where \ }$\mathcal{M}_{\xi }:=${\ }$%
l^{-1}(\xi )${\ {\ is a principal fiber }$G$ {-bundle with the reduced space
\ }$\mathcal{\bar{M}}_{\xi },$ maps {\ $\mathcal{\bar{M}}_{\xi }$\ \
diffeomorphically onto }$T^{\ast }(N).$}
\end{lemma}

Denote by $<.,.>_{\mathcal{G}}$ the standard $Ad$-invariant non-degenerate
scalar product on $\mathcal{G}\times \mathcal{G}.$\ \ The following
characteristic theorem can be derived directly from Lemma \ref{lem_01} .

\begin{theorem}
{\ \label{th_02}\ {Given a principal fiber \ }$G$ {-bundle with a connection 
}$\Gamma (\mathcal{A})$ {\ and a }$G$ {-invariant element }$\ \xi \in 
\mathcal{G}^{\ast },$ {\ then the connection }$\Gamma (\mathcal{A})$ {\
defines a symplectomorphism }$\nu _{\xi }:{\mathcal{\bar{M}}_{\xi }}%
\rightarrow T^{\ast }(N)$ {\ between the reduced phase space }$\mathcal{\bar{%
M}}_{\xi }$ {\ and \ cotangent bundle \ }$T^{\ast }(N),$ {\ where }$%
l:T^{\ast }(M)\rightarrow \mathcal{G}^{\ast }$ {\ is the natural momentum
mapping for the group }$G$ {-action on }$M.$ {\ Moreover, } 
\begin{equation}
(\pi _{\mathcal{A}}^{\xi })(d\text{ }pr_{N}^{\ast }\beta ^{(1)}+pr_{N}^{\ast
}\text{ }\Omega _{\xi }^{(2)})=\left. d\text{ }pr_{M}^{\ast }\alpha
^{(1)}\right\vert _{l^{-1}(\xi )}  \label{0.12}
\end{equation}%
{holds for the canonical 1-forms \ }$\beta ^{(1)}\in \Lambda ^{1}(N)$ {\ and
\ }$\alpha ^{(1)}\in \Lambda ^{1}(M),$ {\ where \ }$\Omega _{\xi
}^{(2)}:=<\Omega ^{(2)},\xi >_{\mathcal{G}}$ {\ is the }$\xi $ {-component
of the corresponding curvature form }$\Omega ^{(2)}\in \Lambda
^{(2)}(N)\otimes \mathcal{G}.$}
\end{theorem}

\begin{remark}
\label{rem_03} {As the canonical 2-form \ }$d\lambda (\alpha ^{(1)})=d$ {\ }$%
pr_{M}^{\ast }\alpha ^{(1)}\in $ {\ }$\Lambda ^{(2)}(T^{\ast }(M))$ {is by
definition }$G$ {-invariant on }$T^{\ast }(M)${, it is evident that its
restriction to the \ }$G$ {-invariant submanifold \ \ \ $\mathcal{M}_{\xi }$}%
$\subset T^{\ast }(M)$ {\ \ will be effectively defined only on the reduced
space $\mathcal{\bar{M}}_{\xi }$} for which{\ (\ref{0.12})is satisfied. \ }
\end{remark}

The following results are direct consequences of Theorem \ref{th_02} that
are useful for many applications \cite{10PS,4Ku}.

\begin{theorem}
\label{th_04} {Let }$\xi \in \mathcal{G}^{\ast }$ {\ have the isotropy group 
}$G_{\xi }$ {acting on the subset \ $\mathcal{M}_{\xi }$}$\subset T^{\ast
}(M)$ {\ freely and properly, so that the reduced phase space \ }$({\mathcal{%
\bar{M}}_{\xi }}\ ,\sigma _{\xi }^{(2)}),$ where ${\mathcal{\bar{M}}_{\xi }:=%
}l^{-1}(\xi )/G_{\xi }$, has symplectic structure defined by{\ } 
\begin{equation}
\sigma _{\xi }^{(2)}:=\left. d\text{ }pr_{M}^{\ast }\alpha ^{(1)}\right\vert
_{l^{-1}(\xi )}.  \label{0.13}
\end{equation}%
{If a principal fiber bundle }$\pi :(M,\varphi )\rightarrow N$ {\ has }$%
G_{\xi }$ as its{\ structure group}$,${\ then\ \ the reduced symplectic
space }$({\mathcal{\bar{M}}_{\xi }},\sigma _{\xi }^{(2)})$ {\ is
symplectomorphic to the cotangent space }$(T^{\ast }(N),\omega _{\xi
}^{(2)}),${\ where } 
\begin{equation}
\omega _{\xi }^{(2)}=d\text{ }pr_{N}^{\ast }\beta ^{(1)}+pr^{\ast }N\Omega
_{\xi }^{(2)},  \label{0.14}
\end{equation}%
{and the corresponding symplectomorphism \ is of the form (\ref{0.12}).}
\end{theorem}

\begin{theorem}
{\label{th_05} In order for two{\ symplectic spaces \ }$({\mathcal{\bar{M}}%
_{\xi }},\sigma _{\xi }^{(2)})$ {\ and \ }$(T^{\ast }(N),dpr_{N}^{\ast
}\beta ^{(1)})$ to be {symplectomorphic, it is necessary and sufficient that
the element \ }$\xi \in \ker ${\ }}${h,}${\ where for the }${G}${{-invariant
element }$\xi \in \mathcal{G}^{\ast }$ {\ the mapping }$h:\xi \rightarrow
\lbrack \Omega _{\xi }^{(2)}]\in H^{2}(N;\mathbb{Z})$, where{\ }$H^{2}(N;%
\mathbb{Z})$ is{\ the cohomology class of 2-forms \ on the manifold }$N.$}
\end{theorem}

\section{Symplectic analysis of Maxwell and Yang-Mills dynamical systems}

Here we shall show how are approach can be applied to various dynamical
systems of the Maxwell and Yang-Mills types.

\subsection{Hamiltonian analysis of Maxwell's electromagnetic dynamical
systems}

We take the Maxwell electromagnetic equations to be%
\begin{eqnarray}
\partial E/\partial t &=&\nabla \times B-J,\text{ \ \ \ }\partial B/\partial
t=-\nabla \times E,  \label{1.1} \\
&<&\nabla ,E>=\rho ,\text{ \ \ \ \ \ \ \ }<\nabla ,B>=0,\text{\ }  \notag
\end{eqnarray}%
on the cotangent phase space $T^{\ast }(N)$, with $N\subset T(D;\mathbb{E}%
^{3})$ - \ the smooth manifold of smooth vector fields on an open domain $%
D\subset \mathbb{R}^{3}$ - all expressed in the light speed units. Here $%
(E,B)\in T^{\ast }(N)$, where the coordinates are the electric and magnetic
fields, respectively, and $\rho :D\rightarrow \mathbb{R}$ and $%
J:D\rightarrow \mathbb{E}^{3}$\ are, respectively, fixed charge density and
current functions on the domain $D,$ satisfying the equation of continuity 
\begin{equation}
\partial \rho /\partial t+<\nabla ,J>=0  \label{1.1a}
\end{equation}%
for all $t\in \mathbb{R}$. Here, $\nabla $ is the gradient operator with
respect to a variable $x\in $ $D,$ $\times $ is the usual vector product in
three-dimensional Euclidean space $\mathbb{E}^{3}:=(\mathbb{R}^{3},<\cdot
,\cdot >)$, which is real three-space $\mathbb{R}^{3}$ endowed with the
usual scalar product $<\cdot ,\cdot >$.

With an eye toward framing equations (\ref{1.1}) in the context of a reduced
symplectic space, we define an appropriate configuration space $M$ $\subset 
\mathcal{T}(D;\mathbb{E}^{3})$ with a vector potential field coordinate \ $%
A\in M.$ The cotangent space $T^{\ast }(M)$\ \ may be identified with pairs $%
(A;Y)\in T^{\ast }(M),$ where $Y\in \mathcal{T}^{\ast }(D;\mathbb{E}^{3})$
is a suitable vector field density in $D.$ There exists the canonical
symplectic form $\omega ^{(2)}\in \Lambda ^{2}(T^{\ast }(M))$ on $T^{\ast
}(M)$ $,$ allowing, $\ $owing to the definition$\ $of\ the\ Liouville\ form$%
\ \ \ \ \ \ \ \ \ \ \ \ \ $%
\begin{equation}
\lambda (\alpha ^{(1)})(A;Y)=\int_{D}d^{3}x(<Y,dA>):=(Y,dA),  \label{1.2}
\end{equation}%
the canonical expression$\ \ \ $%
\begin{equation}
\omega ^{(2)}:=d\lambda (\alpha ^{(1)})=d\text{ }pr_{M}^{\ast }\alpha
^{(1)}=(dY,\wedge dA),  \label{1.2b}
\end{equation}%
where $\wedge $ is the usual exterior Product, $d^{3}x$ denotes Lebesgue
measure in the domain $D$, and $pr_{M}:T^{\ast }(M)\rightarrow M$ \ is the
standard projection upon the base space $M.$ Now we define a Hamiltonian
function $\tilde{H}\in \mathcal{D}(T^{\ast }(M))$ as%
\begin{equation}
\tilde{H}(A,Y)=1/2[(Y,Y)+(\nabla \times A,\nabla \times A)+(<\nabla
,A>,<\nabla ,A>)],  \label{1.2c}
\end{equation}%
to describe the Maxwell equations in vacuo, if the densities $\rho =0$ and $%
J=0.$In fact, owing to (\ref{1.2b}) one easily obtains from (\ref{1.2c})
that 
\begin{eqnarray}
\partial A/\partial t &:&=\delta \tilde{H}/\delta Y=Y,  \label{1.2d} \\
\partial Y/\partial t &:&=-\delta \tilde{H}/\delta A=-\nabla \times B+\nabla
<\nabla ,A>,  \notag
\end{eqnarray}%
which are true wave equations in vacuo, where 
\begin{equation}
B:=\nabla \times A,  \label{1.2e}
\end{equation}%
is the corresponding magnetic field. Now defining 
\begin{equation}
E:=-Y-\nabla W  \label{1.1f}
\end{equation}%
for some function $W:D\rightarrow \mathbb{R}$ as the corresponding electric
field, the system of equations (\ref{1.2d}) assumes, owing to definition (%
\ref{1.2e}), the form 
\begin{equation}
\partial B/\partial t=-\nabla \times E,\text{ \ }\partial E/\partial
t=\nabla \times B,  \label{1.3}
\end{equation}%
which are precisely the Maxwell equations in vacuo, if the Lorentz condition 
\begin{equation}
\partial W/\partial t+<\nabla ,A>=0  \label{1.3a}
\end{equation}%
is imposed.

Since definition (\ref{1.1f}) was essentially imposed rather than arising
naturally from the Hamiltonian approach and our equations are valid only for
a vacuum, we shall try to improve upon these matters by employing the
reduction approach devised in Section 2. Namely, we start with the
Hamiltonian (\ref{1.2c}) and observe that it is invariant with respect to
the abelian symmetry group $G:=\exp \mathcal{G},$ where $\mathcal{G}\simeq
C^{(1)}(D;\mathbb{R}),$ acting on the base manifold $M$ naturally lifted to $%
T^{\ast }(M):$ for any $\psi \in \mathcal{G}$ and $(A,Y)\in T^{\ast }(M)$%
\begin{equation}
\varphi _{\psi }(A):=A+\nabla \psi ,\ \ \ \ \varphi _{\psi }(Y)=Y.
\label{1.4}
\end{equation}%
The 1-form (\ref{1.2}) under the transformation (\ref{1.4}) also is
invariant since 
\begin{equation}
\begin{array}{c}
\varphi _{\psi }^{\ast }\lambda (\alpha ^{(1)})(A,Y)=(Y,dA+\nabla d\psi )=
\\ 
=(Y,dA)-(<\nabla ,Y>,d\psi )=\lambda (\alpha ^{(1)})(A,Y),%
\end{array}
\label{1.5}
\end{equation}%
where we made use of the condition $d\psi \simeq 0$ in $\Lambda ^{1}(T^{\ast
}(M))$ for any $\psi \in \mathcal{G}.$ Thus, the corresponding momentum
mapping (\ref{0.4}) is given as 
\begin{equation}
l(A,Y)=-<\nabla ,Y>  \label{1.6}
\end{equation}%
for all $(A,Y)\in T^{\ast }(M).$ If $\rho \in \mathcal{G}^{\ast }$ is fixed,
one can define the reduced phase space $\mathcal{\bar{M}}_{\rho
}:=l^{-1}(\rho )/G$ since the isotropy group $G_{\rho }=G$, owing to its
commutativity and the condition (\ref{1.4}). Now consider a principal fiber
bundle $\pi :M\rightarrow N$ with the abelian structure group $G$ and a base
manifold $N$ taken as 
\begin{equation}
N:=\{B\in \mathcal{T}(D;\mathbb{E}^{3}):\text{ \ }<\nabla ,\text{ }B>=0,%
\text{ \ }<\nabla ,E(S)>=\rho \},  \label{1.7}
\end{equation}%
where 
\begin{equation}
\pi (A)=B=\nabla \times A.  \label{1.8}
\end{equation}%
We can construct a connection 1-form $\ \ \mathcal{A}\in \Lambda ^{1}(M)%
\mathbb{\otimes }\mathcal{G}$ on this bundle$\mathbf{,}$ such that for all $%
A\in M$, 
\begin{equation}
\mathcal{A}(A)\cdot \hat{A}_{\ast }(l)=1,\text{ \ \ }d<\mathcal{A}(A),\rho
>_{\mathcal{G}}=\Omega _{\rho }^{(2)}(A)\in H^{2}(M;\mathbb{Z}),  \label{1.9}
\end{equation}%
where $\mathcal{A}(A)\in \Lambda ^{1}(M)$ is a differential 1-form, which we
choose as%
\begin{equation}
\mathcal{A}(A):=-(W,d<\nabla ,A>),  \label{1.9a}
\end{equation}%
where $W\in C^{(1)}(D;\mathbb{R})$ is a scalar function, as yet not defined.
As a result, the Liouville form (\ref{1.2}) transforms into 
\begin{equation}
\lambda (\tilde{\alpha}_{\rho }^{(1)}):=(Y,dA)-(W,d<\nabla ,A>)=(Y+\nabla
W,dA):=(\tilde{Y},\text{ }dA),\ \tilde{Y}:=Y+\nabla W,  \label{1.9aa}
\end{equation}%
giving rise to the corresponding canonical symplectic structure on $T^{\ast
}(M)$ as 
\begin{equation}
\tilde{\omega}_{\rho }^{(2)}:=d\lambda (\tilde{\alpha}_{\rho }^{(1)})=(d%
\tilde{Y},\wedge dA).  \label{1.9aaa}
\end{equation}%
Accordingly the Hamiltonian function (\ref{1.2c}), as a function\ on $\
T^{\ast }(M),$ transforms into 
\begin{equation}
\tilde{H}_{\rho }(A,\tilde{Y})=1/2[(\tilde{Y},\tilde{Y})+(\nabla \times
A,\nabla \times A)+(<\nabla ,A>,<\nabla ,A>)],  \label{1.9ab}
\end{equation}%
coinciding with the well-known Dirac-Fock-Podolsky \cite{BS,DFP} Hamiltonian
expression. The corresponding Hamiltonian equations on the cotangent space $%
T^{\ast }(M)$, namely 
\begin{eqnarray*}
\partial A/\partial t &:&=\delta \tilde{H}/\delta \tilde{Y}=\tilde{Y},\text{
\ \ }\tilde{Y}:=-E-\nabla W, \\
\partial \tilde{Y}/\partial t &:&=-\delta \tilde{H}/\delta A=-\nabla \times
(\nabla \times A)+\nabla <\nabla ,A>,
\end{eqnarray*}%
describe true wave processes, related to the Maxwell equations in the vacuo,
except for the external charge and current density conditions. In
particular, from (\ref{1.9ab}) we obtain 
\begin{equation}
\partial ^{2}A/\partial t^{2}-\nabla ^{2}A=0\Longrightarrow \partial
E/\partial t+\nabla (\partial W/\partial t\text{\ }+\text{ }<\nabla
,A>)=-\nabla \times B,\text{\ }  \label{1.9abb}
\end{equation}%
giving rise to the true vector potential wave equation, but the Faraday
induction law is satisfies if one additionally imposes the Lorentz condition
(\ref{1.3a}).

To remedy this situation, we will apply to this symplectic space \ the
reduction technique devised in Section 2. Namely, it follows from Theorem %
\ref{th_04} that above cotangent manifold $T^{\ast }(N)\ $ is
symplectomorphic to the corresponding reduced phase space $\mathcal{\bar{M}}%
_{\rho },$ that is 
\begin{equation}
\mathcal{\bar{M}}_{\rho }\simeq \{(B;S)\in T^{\ast }(N):\ <\nabla
,E(S)>=\rho ,\text{ \ \ }<\nabla ,B>=0\}  \label{1.9b}
\end{equation}%
with the reduced canonical symplectic 2-form 
\begin{equation}
\omega _{\rho }^{(2)}(B,S)=(dB,\wedge dS)=d\lambda (\alpha _{\rho
}^{(1)})(B,S),\text{ \ \ \ }\lambda (\alpha _{\rho }^{(1)})(B,S):=-(S,dB),
\label{1.10}
\end{equation}%
\ where we define

\begin{equation}
\nabla \times S+F+\nabla W=-\tilde{Y}:=E+\nabla W,\text{ \ \ }<\nabla
,F>:=\rho ,  \label{1.10a}
\end{equation}%
for some fixed vector mapping $F\in C^{(1)}(D;\mathbb{E}^{3}),$ depending on
the imposed external charge and current density conditions. The result (\ref%
{1.10}) follows right away upon substituting the expression for the electric
field \ $E=\nabla \times S+F$ into the symplectic structure (\ref{1.9aaa}),
and taking into account the fact that $dF=0$ in $\Lambda ^{1}(M).$Whence,
the Hamiltonian function (\ref{1.9ab}) reduces to the symbolic form%
\begin{eqnarray}
H_{\rho }(B,S) &=&1/2[(B,B)+(\nabla \times S+F+\nabla W,\nabla \times
S+F+\nabla W)+  \notag \\
+( &<&\nabla ,(\nabla \times )^{-1}B>,<\nabla ,(\nabla \times )^{-1}B>)],
\label{1.11}
\end{eqnarray}%
where $"(\nabla \times )^{-1}"$ is the corresponding inverse curl-operation,
mapping \cite{9MW} the divergence-free subspace $C_{\mathrm{div}}^{(1)}(D;%
\mathbb{E}^{3})\subset C^{(1)}(D;\mathbb{E}^{3})$ into itself. Now it
follows from (\ref{1.11}) that \ the Maxwell equations (\ref{1.1}) become a
canonical Hamiltonian system on the reduced phase space $T^{\ast }(N),$
endowed with the canonical symplectic structure (\ref{1.10}) and \ the
modified Hamiltonian function (\ref{1.11}). More precisely, one obtains
easily that%
\begin{eqnarray}
\partial S/\partial t &:&=\delta H/\delta B=B-(\nabla \times )^{-1}\nabla
<\nabla ,(\nabla \times )^{-1}B>,  \label{1.11a} \\
\text{\ \ }\partial B/\partial t &:&=-\delta H/\delta S=-\nabla \times
(\nabla \times S+F+\nabla W)=-\nabla \times E,  \notag
\end{eqnarray}%
where we made use of the definition $E=\nabla \times S+F$ and the elementary
identity $\nabla \times \nabla =0.$ Thus, the second equation of (\ref{1.11a}%
) coincides with \ the second Maxwell equation of (\ref{1.1}) in the
classical form 
\begin{equation*}
\partial B/\partial t=-\nabla \times E.
\end{equation*}%
Moreover, owing to (\ref{1.11a}), from (\ref{1.10a}) one obtains via the
differentiation with respect to $t\in \mathbb{R}$ that 
\begin{eqnarray}
\partial E/\partial t &=&\partial F/\partial t+\nabla \times \partial
S/\partial t=  \label{1.11b} \\
&=&\partial F/\partial t+\nabla \times B,  \notag
\end{eqnarray}%
as well as, owing to (\ref{1.1a}), 
\begin{equation}
<\nabla ,\partial F/\partial t>=\partial \rho /\partial t=-<\nabla ,J>.
\label{1.11c}
\end{equation}%
Now we can write down from (\ref{1.11c}) that, up to non-essential
curl-terms $\nabla \times (\cdot ),$ the following relationship%
\begin{equation}
\partial F/\partial t=-J  \label{1.11d}
\end{equation}%
holds. In fact, the current vector $J\in C^{(1)}(D;\mathbb{E}^{3}),$ owing
to the equation of continuity (\ref{1.1a}), is defined up to curl-terms $%
\nabla \times (\cdot )$ which can be included in the definition of the
right-hand side of (\ref{1.11d}). Then upon substitution of (\ref{1.11d})
into (\ref{1.11b}), we obtain the first Maxwell equation of (\ref{1.1}): 
\begin{equation}
\partial E/\partial t=\nabla \times B-J,  \label{1.11cd}
\end{equation}%
which is naturally supplemented with the external charge and current
densities conditions 
\begin{equation}
\begin{array}{c}
<\nabla ,B>=0,\text{ \ \ }<\nabla ,E>=\rho , \\ 
\partial \rho /\partial t+<\nabla ,J>=0,%
\end{array}
\label{1.11cdd}
\end{equation}%
in virtue of the equation of continuity (\ref{1.1a}) and definition (\ref%
{1.9b}).

As for the wave equations related to the Hamiltonian system (\ref{1.11a}), \
we find that the electric field $E$ is recovered from the second equation as 
\begin{equation}
E:=-\partial A/\partial t-\nabla W,  \label{1.11dd}
\end{equation}%
where $W\in C^{(1)}(D;\mathbb{R})$ is a smooth function that depends on the
vector field $A\in M.$ To determine this dependence, we substitute (\ref%
{1.11d}) \ into equation (\ref{1.11cd}) taking into account that $B=\nabla
\times A$, which yields%
\begin{equation}
\partial ^{2}A/\partial t^{2}-\nabla (\partial W/\partial t+<\nabla
,A>)=\nabla ^{2}A+J.  \label{1.11de}
\end{equation}%
\ With the above, if we now impose the Lorentz condition (\ref{1.3a}), we
obtain from (\ref{1.11de}) \ the corresponding true wave equations in the
space-time, taking into account the external charge and current density
conditions (\ref{1.11cdd}).

Notwithstanding our progress so far, the problem of fulfilling the Lorentz
constraint (\ref{1.3a}) naturally within the canonical Hamiltonian
formalism\ still remains to be completely solved. To this end, we are
compelled to analyze the structure of the \ Liouville 1-form (\ref{1.9aa})
for the Maxwell equations on a slightly extended functional manifold $%
M\times L.$ As the first step, we rewrite the 1-from (\ref{1.9aa}) as 
\begin{eqnarray}
\lambda (\tilde{\alpha}_{\rho }^{(1)}) &:&=(\tilde{Y},dA)=(Y+\nabla
W,dA)=(Y,dA)+  \notag \\
+(W,-d &<&\nabla ,A>):=(Y,dA)+(W,d\eta ),  \label{1.11e}
\end{eqnarray}%
where 
\begin{equation}
\eta :=-<\nabla ,A>.  \label{1.11f}
\end{equation}%
Considering now the elements $(Y,A;\eta ,W)$ $\in T^{\ast }(M\times L)$ as
new independent canonical variables on the extended cotangent phase space $%
T^{\ast }(M\times L),$ where $L:=C^{(1)}(D;\mathbb{R}),$ we can rewrite the
symplectic structure (\ref{1.9aaa}) in the following canonical form%
\begin{equation}
\tilde{\omega}_{\rho }^{(2)}:=d\lambda (\tilde{\alpha}_{\rho
}^{(1)})=(dY,\wedge dA)+(dW,\wedge d\eta ).  \label{1.11g}
\end{equation}%
In view of the Hamiltonian function (\ref{1.9ab}), we obtain the expression 
\begin{equation}
H(A,Y;\eta ,W)=1/2[(Y-\nabla W,Y-\nabla W)+(\nabla \times A,\nabla \times
A)+(\eta ,\eta )],  \label{1.11h}
\end{equation}%
with respect to which the corresponding Hamiltonian equations take the form%
\begin{eqnarray}
\partial A/\partial t &:&=\delta H/\delta Y=Y-\nabla W,\text{ \ \ }Y:=-E, 
\notag \\
\partial Y/\partial t &:&=-\delta H/\delta A=-\nabla \times (\nabla \times
A),  \notag \\
\partial \eta /\partial t &:&=\delta H/\delta W=<\nabla ,Y-\nabla W>,  \notag
\\
\partial W/\partial t &:&=-\delta H/\delta \eta =-\eta .  \label{1.11hh}
\end{eqnarray}%
From (\ref{1.11hh}), we readily compute that%
\begin{eqnarray}
\partial B/\partial t+\nabla \times E &=&0,\text{ \ }\partial ^{2}W/\partial
t^{2}-\nabla ^{2}W=<\nabla ,E>,  \label{1.11hi} \\
\partial E/\partial t-\nabla \times B &=&0,\text{ \ \ }\partial
^{2}A/\partial t^{2}-\nabla ^{2}A=-\nabla (\partial W/\partial t+<\nabla
,A>).  \notag
\end{eqnarray}%
It is evident that these equations describe Maxwell's equations in the
vacuo, without taking into account both the external charge and current
density relationships (\ref{1.11cdd}) and the Lorentz condition (\ref{1.3a}%
). Our next step is to apply the reduction technique devised in Section 2 to
the symplectic structure (\ref{1.11g}) . Whence we find that under the
transformations (\ref{1.10a}), the corresponding reduced manifold $\mathcal{%
\bar{M}}_{\rho }$ becomes endowed with the symplectic structure 
\begin{equation}
\bar{\omega}_{\rho }^{(2)}:=(dB,\wedge dS)+(dW,\wedge d\eta ),  \label{1.11i}
\end{equation}%
and the Hamiltonian (\ref{1.11h}) assumes the form 
\begin{equation}
H(S,B;\eta ,W)=1/2[(\nabla \times S+F+\nabla W,\nabla \times S+F+\nabla
W)+(B,B)+(\eta ,\eta )].  \label{1.11j}
\end{equation}%
The Hamiltonian equations for $H$ are 
\begin{eqnarray}
\partial S/\partial t &:&=\delta H/\delta B=B,\text{ \ \ \ \ \ \ }\partial
W/\partial t:=-\delta H/\delta \eta =-\eta ,  \label{1.11k} \\
\partial B/\partial t &:&=-\delta H/\delta S=-\nabla \times (\nabla \times
S+F+\nabla W)=-\nabla \times E,  \notag \\
\partial \eta /\partial t &:&=\delta H/\delta W=-<\nabla ,\nabla \times
S+F+\nabla W>=-<\nabla ,E>-\Delta W,  \notag
\end{eqnarray}%
which coincide under the constraint (\ref{1.10a}) completely with Maxwell
equations (\ref{1.1}), describing true space-time processes and taking into
account, \textit{a priori,} both the imposed external charge and current
density relationships (\ref{1.11cdd}) and the Lorentz condition (\ref{1.3a}%
),thus solving the problem mentioned in \cite{BS,DFP}. Indeed, it is easy to
obtain from (\ref{1.11k}) that 
\begin{eqnarray}
\partial ^{2}W/\partial t^{2}-\Delta W &=&\rho ,\text{ \ \ \ \ \ \ \ \ \ }%
\partial W/\partial t+<\nabla ,A>=0,  \label{1.11l} \\
\nabla \times B &=&J+\partial E/\partial t,\text{ \ \ \ \ \ \ \ }\partial
B/\partial t=-\nabla \times E,  \notag
\end{eqnarray}%
Hence, using (\ref{1.11l}) and (\ref{1.11cdd}), one can easily calculate 
\cite{BP,BPT} the magnetic wave equation%
\begin{equation}
\partial ^{2}A/\partial t^{2}-\Delta A=J,  \label{1.11m}
\end{equation}%
supplementing the suitable wave equation on the scalar potential $W\in L,$
thereby completing the calculations. Thus, we have proved the desired
result; namely,

\begin{proposition}
The electromagnetic Maxwell equations (\ref{1.1}) together with Lorentz
condition (\ref{1.3a}) are equivalent to the Hamiltonian system (\ref{1.11k}%
) \ with respect to the canonical symplectic structure \ (\ref{1.11i}) \ and
Hamiltonian function (\ref{1.11j}), which, respectively, reduce to the
electromagnetic equations (\ref{1.11l}) and (\ref{1.11m}) under the external
charge and current density relationships (\ref{1.11cdd}).
\end{proposition}

The above result can be used\ for developing an \ alternative quantization
procedure\ of Maxwell's \ equations, as it circumvents the related quantum
operator compatibility problems discussed in detail in \cite{BS,BjD,DFP}. We
hope to consider this aspect of the quantization problem in a future
investigation.

\begin{remark}
{\ If one to considers the motion of a charged point particle under a
Maxwell field, it is convenient to introduce a trivial fiber bundle
structure }${\pi }${$:M\rightarrow N,$ \ such that $M=N\times G$}${,}${\ $\
N:=D\subset \mathbb{R}^{3}$ and $G:=\mathbb{R}/\{0\}{\ }$is the
corresponding (abelian) structure Lie group. An analysis similar to the
above gives rise to the reduced (on the space }$\mathcal{\bar{M}}_{\xi }{:=l}
${$^{-1}(\xi )/G\simeq T^{\ast }(N),$ $\xi \in \mathcal{G}$)\ symplectic
structure }%
\begin{equation*}
{\omega ^{(2)}(q,p)=<dp,\wedge dq>+d<\mathcal{A}(q,g),\xi >_{\mathcal{G}},}
\end{equation*}%
{\ where $\mathcal{A}(q,g):=<A(q),dq>+g^{-1}dg$ is a suitable connection
1-form on the phase\ space $\ M,$ with $(q,p)\in T^{\ast }(N)$ and $g\in G.$
The corresponding canonical Poisson brackets on $T^{\ast }(N)$ are easily
found to be 
\begin{equation}
\{q^{i},q^{j}\}=0,\text{ \ \ }\{p_{j},q^{i}\}=\delta _{j}^{i},\text{ \ \ \ \
\ \ }\{p_{i},p_{j}\}=F_{ji}(q)  \label{1.21}
\end{equation}%
for all $(q,p)\in T^{\ast }(N).$ If one introduces a new momentum variable $%
\tilde{p}:=p+A(q)$ on $T^{\ast }(N)\ni (q,p),$ it is easy to verify that \ $%
\omega _{\xi }^{(2)}\rightarrow \tilde{\omega}_{\xi }^{(2)}:=<d\tilde{p}%
,\wedge dq>$}${,}${\ which gives rise to the following Poisson brackets \cite%
{8Kup,10Pr,10PS}: 
\begin{equation}
\{q^{i},q^{j}\}=0,\text{ \ \ \ \ }\{\tilde{p}_{j},q^{i}\}=\delta _{j}^{i},%
\text{ \ \ \ \ \ \ }\{\tilde{p}_{i},\tilde{p}_{j}\}=0,  \label{1.22}
\end{equation}%
where $i,j=\overline{1,3},$ iff \ for all $i,j,k=\overline{1,3}$ \ the
standard Maxwell field equations are satisfied on $N:$%
\begin{equation}
\partial F_{ij}/\partial q_{k}+\partial F_{jk}/\partial q_{i}+\partial
F_{ki}/\partial q_{j}=0  \label{1.23}
\end{equation}%
with the curvature tensor $F_{ij}(q):=\partial A_{j}/\partial q^{i}-\partial
A_{i}/\partial q^{j},$ \ $i,j=\overline{1,3},$ $q\in N.$}
\end{remark}

It is not difficult to see that the above approach permits a natural
generalization for non-abelian structure Lie groups, yielding a description
of Yang-Mills field equations within our reduction formulation. We proceed
to such an extension in the next subsection.

\subsection{Hamiltonian analysis of Yang-Mills dynamical systems}

As above, we start by defining a phase space $M$ of a particle moving under
a Yang-Mills field in a region $D\subset \mathbb{R}^{3}$ with $M:=D\mathbb{%
\times }G,$ where $G$ is a (not in general semisimple) Lie group, acting on $%
M$ \ from the right. Over the space $M$ one can define quite naturally a
connection $\Gamma (\mathcal{A})$ by consider the trivial principal fiber
bundle $\pi :M\rightarrow N,$ where $N:=D,$ with the structure group $G.$
Namely, if $g\in G,$ $\ q\in N,$ then a connection 1-form on $M\ni (q,g)$
can be expressed \cite{1GS,3PM,3HPP,7Mo} as 
\begin{equation}
\mathcal{A}(q;g):=g^{-1}(d+\sum_{i=1}^{n}a_{i}A^{(i)}(q))g,  \label{2.1}
\end{equation}%
where $\{a_{i}\in \mathcal{G}:i=\overline{1,n}\}$ is a basis for the Lie
algebra $\mathcal{G}$ \ of the Lie group $G$, and $A_{i}:D\rightarrow
\Lambda ^{1}(D),$ $i=\overline{1,n},$ are the Yang-Mills fields on the
physical space $D\subset \mathbb{R}^{3}.$

Now one defines the natural left invariant Liouville form on $M$ \ as 
\begin{equation}
\lambda (\alpha ^{(1)})(q;g):=<p,dq>+<y,g^{-1}dg>_{\mathcal{G}},  \label{2.2}
\end{equation}%
where $y\in T^{\ast }(G)$ and $\ <\cdot ,\cdot >_{\mathcal{G}}$ denotes as
before the usual Ad-invariant nondegenerate bilinear form on $\mathcal{G}%
^{\ast }\times \mathcal{G},$ and it is clear that $g^{-1}dg\in \Lambda
^{1}(G)\otimes \mathcal{G}\mathbf{.}$ The main assumption we need to proceed
is that the connection 1-form is compatible with the Lie group $G$ \ action
on $M.$ The means that 
\begin{equation}
R_{h}^{\ast }\mathcal{A}(q;g)=Ad_{h^{-1}}\mathcal{A}(q;g)  \label{2.3}
\end{equation}%
is satisfied for all \ $(q,g)\in M$ and $h\in G,$ where $R_{h}:G\rightarrow
G $ is the right translation by an element $h\in G$ on the Lie group $\ G.$

Having gathered all preliminary elements needed for the reduction Theorem %
\ref{th_04} to be applied to our model, we now suppose that the Lie group $G$
canonical action on $M$ is naturally lifted to the cotangent space $T^{\ast
}(M)$ endowed, owing to ({\ref{1.2}), with the $G$-invariant canonical
symplectic structure \ \ \ \ 
\begin{eqnarray}
\omega ^{(2)}(q,p;g,y) &:&=d\text{ }pr_{M}^{\ast }\alpha
^{(1)}(q,p;g,y)=<dp,\wedge dq>+  \label{2.4} \\
+ &<&dy,\wedge g^{-1}dg>_{\mathcal{G}}+<ydg^{-1},\wedge dg>_{\mathcal{G}} 
\notag
\end{eqnarray}%
for all $(q,p;g,y)\in T^{\ast }(M).$ Choose an element $\xi \in \mathcal{G}%
^{\ast }$ and assume that its isotropy subgroup $G_{\xi }=G,$ that is $%
Ad_{h}^{\ast }\xi =\xi $ for all $h\in G.$ In the general case such an
element $\xi \in \mathcal{G}^{\ast }$ cannot exist unless it is trivial, $%
\xi =0,$ as it happens, for instance, in the case of the Lie group $G=SL_{2}(%
\mathbb{R}).$ Then one can construct the reduced phase space $l^{-1}(\xi )/G$
symplectomorphic to $(T^{\ast }(N),\omega _{\xi }^{(2)}),$ where it follows
from (\ref{0.12}) that for any $(q,p)\in T^{\ast }(N)$,%
\begin{eqnarray}
\omega _{\xi }^{(2)}(q,p) &=&<dp,\wedge dq>+<\Omega ^{(2)}(q),\xi >_{%
\mathcal{G}}=  \label{2.5} \\
&=&<dp,\wedge
dq>+\sum_{s=1}^{n}\sum_{i,j=1}^{3}e_{s}F_{ij}^{(s)}(q)dq^{i}\wedge dq^{j}. 
\notag
\end{eqnarray}%
In the above we have expanded the element $\xi =\sum_{i=1}^{n}e_{i}a^{i}$ }$%
\in \mathcal{G}^{\ast }${{\ }with respect to the bi-orthogonal basis $%
\{a^{i}\in \mathcal{G}^{\ast },a_{j}\in \mathcal{G}:$ $\ <a^{i},a_{j}>_{%
\mathcal{G}}=\delta _{j}^{i},$ $i,j=\overline{1,n}\},$ \ with constant
coefficients $e_{i}\in \mathbb{R},$ $i=\overline{1,3}$. We also denoted by $%
F_{ij}^{(s)}(q),$ $i,j=\overline{1,3},$ $s=\overline{1,n},$ the
corresponding curvature 2-form $\Omega ^{(2)}\in \Lambda ^{2}(N)\otimes 
\mathcal{G}$ components, that is 
\begin{equation}
\Omega ^{(2)}(q):=\sum_{s=1}^{n}\sum_{i,j=1}^{3}a_{s\text{ }%
}F_{ij}^{(s)}(q)dq^{i}\wedge dq^{j}  \label{2.6}
\end{equation}%
for any point $q\in N.$ Summarizing the calculations above, we have the
following result. }

\begin{theorem}
\label{th_2.1} {\ Suppose the Yang-Mills field (\ref{2.1}) on the fiber
bundle }$\pi :M\rightarrow N$ {\ with }$M=D\times G$ {\ is invariant with
respect to the Lie group }$G$ {\ action }$G\times M\rightarrow M.$ {\
Suppose also that an element }$\xi \in G^{\ast }$ {\ is chosen so that }$%
Ad_{G}^{\ast }\xi =\xi .$ {\ Then for the naturally constructed momentum
mapping }$l:T^{\ast }(M)\rightarrow G^{\ast }$ {\ (which is equivariant),
the reduced phase space }$l^{-1}(\xi )/G\simeq T^{\ast }(N)$ {\ is endowed
with the symplectic structure (\ref{2.5}), having the component-wise Poisson
brackets form} 
\begin{equation}
\{p_{i},q^{j}\}_{\xi }=\delta _{i}^{j},\text{ \ \ }\{q^{i},q^{j}\}_{\xi
}=0,\ \ \{p_{i},p_{j}\}_{\xi }=\sum_{s=1}^{n}e_{s}F_{ji}^{(s)}(q)
\label{2.7}
\end{equation}%
{for all }$i,j=\overline{1,3}$ {\ and }$(q,p)\in T^{\ast }(N).$
\end{theorem}

The corresponding extended Poisson bracket on the whole cotangent space $%
T^{\ast }(M)$ comprises, owing to (\ref{1.4}), the following set of Poisson
relationships: 
\begin{eqnarray}
\{y_{s},y_{k}\}_{\xi } &=&\sum_{r=1}^{n}c_{sk\text{ }}^{r}y_{r},\text{ \ \ \
\ \ \ \ \ \ \ \ \ }\ \ \{p_{i},q^{j}\}_{\xi }=\ \delta _{i}^{j}\ \ ,\text{ }
\label{2.8} \\
\text{\ }\{y_{s},p_{j}\}_{\xi } &=&0=\{q^{i},q^{j}\},\text{\ \ }%
\{p_{i},p_{j}\}_{\xi }=\sum_{s=1}^{n}y_{s\text{ }}F_{ji}^{(s)}(q),  \notag
\end{eqnarray}%
where $i,j=\overline{1,n},$ $\ c_{sk}^{r}\in \mathbb{R},$ \ $s,k,r=\overline{%
1,m},$ are the structure constants of the Lie algebra $\mathcal{G},$ and we
made use of the expansion \ $A^{(s)}(q)=\sum_{j=1}^{n}A_{j}^{(s)}(q)$ $%
dq^{j} $ as well as introducing alternative fixed values $e_{i}:=y_{i},$ $i=%
\overline{1,n}.$ The result (\ref{2.8}) follows readily by making the shift
in the expression (\ref{2.4}) defined as $\sigma ^{(2)}\rightarrow \sigma
_{ext}^{(2)},$ where $\sigma _{ext}^{(2)}:=\left. \sigma ^{(2)}\right\vert _{%
\mathcal{A}_{0}\rightarrow \mathcal{A}}$ $,$ $\mathcal{A}_{0}(g):=g^{-1}dg,$ 
$g\in G.$ With this, the invariance properties of the connection $\Gamma (%
\mathcal{A})$ imply that 
\begin{equation*}
\sigma _{ext}^{(2)}(q,p;u,y)=<dp,\wedge dq>+d<y(g),Ad_{g^{-1}}\mathcal{A}%
(q;e)>_{\mathcal{G}}=
\end{equation*}%
\begin{equation*}
=<dp,\wedge dq>+<d\text{ }Ad_{g^{-1}}^{\ast }y(g),\wedge \mathcal{A}(q;e)>_{%
\mathcal{G}}=<dp,\wedge dq>+\sum_{s=1}^{m}dy_{s}\wedge du^{s}+
\end{equation*}

\begin{equation*}
+\sum_{j=1}^{n}\sum_{s=1}^{m}A_{j}^{(s)}(q)dy_{s}\wedge
dq-<Ad_{g^{-1}}^{\ast }y(g),\mathcal{A}(q,e)\wedge \mathcal{A}(q,e)>_{%
\mathcal{G}}+
\end{equation*}%
\begin{equation}
+\sum_{k\geq s=1}^{m}\sum_{l=1}^{m}y_{l}\text{ }c_{sk}^{l}\text{ }%
du^{k}\wedge du^{s}+\sum_{s=1}^{n}\sum_{i\geq
j=1}^{3}y_{s}F_{ij}^{(s)}(q)dq^{i}\wedge dq^{j},  \label{2.9}
\end{equation}%
where the coordinates of $(q,p;u,y)\in T^{\ast }(M)$ \ are defined as
follows: $\mathcal{A}_{0}(e):=\sum_{s=1}^{m}du^{i}$ $a_{i},$ and $%
Ad_{g^{-1}}^{\ast }y(g)=y(e):=\sum_{s=1}^{m}y_{s}$ $a^{s}$ for any element $%
g\in G.$ This leads immediately to the Poisson brackets (2.8) plus
additional brackets connected with conjugated sets of variables $\{u^{s}\in 
\mathbb{R}:$\texttt{\ }$s=\overline{1,m}\}$ $\in \mathcal{G}^{\ast }$\ and $%
\{y_{s}\in \mathbb{R}:$\texttt{\ }$s=\overline{1,m}\}\in \mathcal{G}:$

\begin{equation}
\{y_{s},u^{k}\}_{\xi }=\delta _{s}^{k},\text{ \ }\{u^{k},q^{j}\}_{\xi }=0,%
\text{ \ }\{p_{j},u^{s}\}_{\xi }=A_{j}^{(s)}(q),\text{ \ }%
\{u^{s},u^{k}\}_{\xi }=0,  \label{2.10}
\end{equation}%
where $j=\overline{1,n},$ \ $k,s=\overline{1,m},\ $and $\ \ q\in N.$

Note here that the transition from the symplectic structure $\sigma ^{(2)}$
\ on $T^{\ast }(N)$ to its extension $\sigma _{ext}^{(2)}$ on \ $T^{\ast
}(M) $ suggested above just consists formally in adding an exact part to the
symplectic structure $\sigma ^{(2)}$, which transforms it into equivalent
one. Looking now at the expressions (\ref{2.9}), one can infer immediately
that an element \ $\xi :=\sum_{s=1}^{m}e_{s}a^{s}\in \mathcal{G}^{\ast }$ \
will be invariant with respect to the $Ad^{\ast }$-action of the Lie group $%
\ G$ \ iff 
\begin{equation}
\left. \{y_{s},y_{k}\}_{\xi }\right\vert
_{y_{s}=e_{s}}=\sum_{r=1}^{m}c_{sk}^{r}\text{ }e_{r}\text{ }=0  \label{2.11}
\end{equation}%
identically for all $s,k=\overline{1,m},$ \ $j=\overline{1,n}$ \ and $\ q\in
N.$ In this and only this case does the reduction scheme elaborated above go
through.

Returning our attention to the expression (\ref{2.10}), one can easily
derive the exact shifted expression 
\begin{equation}
\omega _{ext}^{(2)}(q,p;u,y)=\omega ^{(2)}(q,p+\sum_{s=1}^{n}y_{s}\text{ }%
A^{(s)}(q)\text{ };u,y),  \label{2.12}
\end{equation}%
on the phase space $T^{\ast }(M)\ni (q,p;u,y),$ where we abbreviated for
brevity $<A^{(s)}(q),dq>$ as $\sum_{j=1}^{n}A_{j}^{(s)}(q)$ $dq^{j}.$
Expressions like (\ref{2.12}) were discussed within a somewhat different
context in \cite{8Kup,10Pr}, which also provide a good background for the
infinite-dimensional generalization of the symplectic structure techniques.
Having observed from (\ref{2.12}) that the simple change of variables 
\begin{equation}
\tilde{p}:=p+\sum_{s=1}^{m}y_{s}\text{ }A^{(s)}(q)  \label{2.13}
\end{equation}%
in the cotangent space $T^{\ast }(N)$ recasts our symplectic structure (\ref%
{2.9}) into the old canonical form (\ref{2.4}), one obtains that the
following new set of canonical Poisson brackets on $T^{\ast }(M)$ $\ni (q,%
\tilde{p};u,y):$%
\begin{eqnarray}
\{y_{s},y_{k}\}_{\xi } &=&\sum_{r=1}^{n}c_{sk}^{r}\text{ }y_{r},\text{ \ \ \ 
}\{\tilde{p}_{i},\tilde{p}_{j}\}_{\xi }=0,\text{\ \ \ \ \ }\{\tilde{p}%
_{i},q^{j}\}=\delta _{i}^{j},\text{ }  \label{2.14} \\
\{y_{s},q^{j}\}_{\xi } &=&0\text{ }=\{q^{i},q^{j}\}_{\xi },\text{\ }%
\{u^{s},u^{k}\}_{\xi }=0,\text{ \ \ }\{y_{s},\tilde{p}_{j}\}_{\xi }=0,\text{
\ }  \notag \\
\{u^{s},q^{i}\}_{\xi } &=&0,\text{ \ \ \ \ \ \ \ \ \ }\{y_{s},u^{k}\}_{\xi
}=\delta _{s}^{k},\text{ \ \ \ \ \ \ \ \ \ \ }\{u^{s},\tilde{p}_{j}\}_{\xi
}=0,  \notag
\end{eqnarray}%
where $\ k,s=\overline{1,m}$ \ and $i,j=\overline{1,n},$ holds iff the
nonabelian Yang-Mills field equations 
\begin{equation}
\partial F_{ij}^{(s)}/\partial q^{l}+\partial F_{jl}^{(s)}/\partial
q^{i}+\partial F_{li}^{(s)}/\partial q^{j}+  \label{2.15}
\end{equation}%
\begin{equation*}
+\sum_{k,r=1}^{m}c_{kr}^{s}(F_{ij}^{(k)}A_{l}^{(r)}+F_{jl}^{(k)}A_{i}^{(r)}+F_{li}^{(k)}A_{j}^{(r)})=0
\end{equation*}%
are fulfilled for all $\ s=\overline{1,m}$ \ and $i,j,l=\overline{1,n}$ on
the base manifold $\ N.$ This effect of complete reduction of gauge
Yang-Mills variables from the symplectic structure (\ref{2.9}) is known in
literature \cite{8Kup} as the principle of minimal interaction and has
proven to be quite useful for studying different interacting systems as in 
\cite{9MW,10PZ}. We plan to continue the study of the geometric properties
of reduced symplectic structures connected with such interesting
infinite-dimensional coupled dynamical systems as those of
Yang-Mills-Vlasov, Yang-Mills-Bogolubov and Yang-Mills-Josephson types \cite%
{9MW,10PZ}, as well as their relationships with associated principal fiber
bundles endowed with canonical connection structures.

\section{Acknowledgments}

The authors are cordially thankful to the Abdus Salam International Centre
for Theoretical Physics in Trieste, Italy, for the hospitality during their
2007-2008 research scholarships. A.P. is especially grateful to Profs. P.I.
Holod (Kyiv, UKMA), J.M. Stakhira and I.M. Bolesta (Lviv, LNU) \ for
fruitful discussions, useful comments and remarks. The authors also express
their sincere appreciation to the referees whose insightful remarks,
suggestions and valuable comments were instrumental to improving the
manuscript. Last but not least, thanks go to Prof. D.L. Blackmore (NJIT) for
his help in editing the article and academician Prof. A.A. Logunov (Moscow,
IJP) for his interest in the work.

\end{document}